\documentclass[
 reprint,
superscriptaddress,
aps,
 amsmath,amssymb,
pra,
]{revtex4-1}

\usepackage{graphics}
\usepackage{graphicx}
\usepackage{dcolumn}
\usepackage{bm}

\begin{document}

\title{Quantum criticality analysis by finite size scaling and exponential basis sets}%

\author{Fahhad H. Alharbi}%
\email{falharbi@qf.org.qa}
\affiliation{Qatar Environment and Energy Research Institute (QEERI), Doha, Qatar}
\affiliation{King Abdulaziz City for Science and Technology (KACST), Riyadh, Saudi Arabia}

\author{Sabre Kais}%
\email{kais@purdue.edu}
\affiliation{Qatar Environment and Energy Research Institute (QEERI), Doha, Qatar}
\affiliation{Department of Chemistry, Physics, and Birck Nanotechnology Center, Purdue University, West Lafayette,
Indiana 47907, USA)}

\begin{abstract}

We combine the finite size scaling method with the meshfree spectral method  to calculate quantum critical parameters for a given Hamiltonian. The basic idea is to expand the exact  wave function in a finite exponential basis set and extrapolate  the information about system criticality from a finite basis to  the infinite basis set limit. The used exponential basis set -though chosen intuitively- allows handling a very wide range of exponential decay rates and calculating multiple eigenvalues simultaneously. As a benchmark system to illustrate the combined approach, we choose the  Hulthen potential. The results show that the method is very accurate and converges faster when compared with other basis functions. The approach is general and can be extended to examine near threshold phenomena for atomic and molecular systems based on even-tempered exponential and Gaussian basis functions. 

\end{abstract}
\maketitle

\section{Introduction}

The study of how the energy levels of a given system  change as one varies a parameter in the corresponding Hamiltonian is of general interest, particularly near binding threshold, level crossings and quantum phase transitions. In phase transitions, critical points are associated with singularities of the free energy which occur only
in the thermodynamic limit \cite{Y01,Y02}. Finite size scaling (FSS) was developed by Fisher and others \cite{F01,N01,D01,P01,S01} to calculate such parameters by extrapolating information from a finite system to the thermodynamic limit. In analogy, FSS was also developed to extrapolate information from a finite basis set to the infinite basis set limit in order to calculate quantum critical parameters for a given Hamiltonian. This is done by expanding the exact wave function in a complete basis set and use the number of basis function to play the role of system size \cite{Kais-Serra}. Early work using FSS to calculate quantum critical parameters was based on expanding the wave function in Slater-type and Gaussian-type functions \cite{S02,S03,K01}. Recently, the method was also combined with the finite element method (FEM) \cite{M01,A01} and B-splines expansion to achieve similar results \cite{Serra-2012}.

Here, we combine FSS method with the meshfree spectral method (SM) to calculate quantum critical parameters. Lately, the meshfree SMs start gaining growing attention because of their high levels of analyticity and accuracy \cite{B01,C01,G01,FH03,FH04}. In these methods, the unknown functions are approximated by expansion using preselected   basis sets. One of the main challenges in SM is to handle domains extended to infinity \cite{F02,C02,G02,G03,V01,K02,S04}. Many techniques were introduced to overcome this challenge such as using exponentially decaying functions as basis sets, the truncation of the computational windows, and applying size scaling. Recently, a non-orthogonal predefined exponential basis set for eignevalue problems involving half bounded domains was introduced and used \cite{FH01,FH02}. The set is easy to use and allows generally finding a wide range of eigenvalues simultaneously.

In this paper, the exponential basis sets are implemented in FSS analysis to obtain the quantum critical parameters for a given  Hamiltonian. The presented technique is real-space meshfree. Such real-space techniques start gaining more attentions in "ab initio" and density functional calculations \cite{B04,M01}. As a benchmark system we choose the  Hulthen potential.  For such Hamiltonian, the analytical solution is known and also FSS was implemented using other basis functions, hence our numerical results can be compared and analyzed. The comparison confirms the validity and efficiency of the new approach and its applicability for FSS analysis which will be used on more complex systems.

\section{Theoretical background}

\subsection{Analytical solution for the Hamiltonian with Hulthen potential}

Hulthen potential \cite{H01,H02} is a special case of Eckart potential \cite{E01}, which is a family of screened Coulomb potentials. It has the following form
\begin{equation}
	\label{Hulthen}
	V(r) = - \frac{\lambda}{a^2} \frac{e^{-r/a}}{1-e^{-r/a}}
\end{equation}
where $\lambda$ is the coupling constant and $a$ is the scaling parameter. For small $r$, it resembles Coulomb potential. But, it dies faster and exponentially for large $r$. 

By defining a dimensionless variable, $ x = \frac{r}{a}$, 
and inserting the potential in Schr\"{o}dinger radial differential equation, the radial equation becomes
\begin{equation}
	\label{RWE1}
	\left[ - \frac{1}{2}\frac{d^2}{dx^2} - \lambda \frac{e^{-x}}{1-e^{-x}} \right] \psi = a^2 E \psi
\end{equation}
The analytical solution for this equation is known in term of hypergeometric function and it is
\begin{multline}
\label{ASol}
\psi = N_0 e^{-\overline{a} x} \left( 1 - e^{-x} \right) \\ {}_
2F_1 \left( 2\overline{a}+1+n, 1-n, 2\overline{a}+1; e^{-x} \right)
\end{multline}
where $\overline{a} = - a^2 E$, $n$ is the state order, and $N_0$ is the normalization factor and it is
\begin{equation}
	\label{Normalization}
	N_0 = \sqrt{\overline{a} \left( \overline{a} + n \right) \left( 2 \overline{a} + n \right)} \frac{\Gamma \left( 2\overline{a}+n \right)}{\Gamma \left( 2\overline{a}+1 \right) \Gamma \left( n \right)}
\end{equation}
The energy levels are
\begin{multline}
	\label{Levels}
	E_n = - \frac{1}{a^2} \frac{\left( 2\lambda - n^2 \right)^2}{8 n^2},  \\
	\qquad for \qquad n = 1,2,3,\cdots, n_{max}
\end{multline}

It is clear from Eq.(\ref{Levels}) that $\lambda_c = n^2/2$ is a critical coupling constant. As $\lambda_c$ is a function of $n$, it is obvious that the number of allowed bounded states (i.e. $n_{max}$) is $\lambda$ dependent. It has at least one state for $\lambda>1/2$ and this is the critical point to be tracked.

\subsection{Finite size scaling}

As aforementioned, FSS method is a systematic approach allowing extrapolating the critical behaviour of an infinite system by analysing a finite sample of it. It is efficient and accurate for the calculation of critical parameters of the Schrödinger equation. Assuming that the Hamiltonian of a system is of the following form
\begin{equation}
	\label{HamForm}
	H = H_0 + V_\lambda \left( \lambda \right)
\end{equation}
where again $\lambda$ is the coupling constant.  The critical point,  $\lambda_c$, will be defined as a point for which a bound state becomes absorbed or
degenerate with a continuum.

As known, the asymptotic behaviours of physical quantities near to the critical points are associated with critical exponents. So, the energy near to $\lambda_c$ can be defined as
\begin{equation}
	\label{ExpForm}
	E_\lambda- E_{th} = \left( \lambda - \lambda_c \right)^\alpha,
\end{equation}
where we assume that the threshold energy, $E_{th}$ does not depend on $\lambda$.
In principle, $\lambda_c$ can be calculated providing the exact solution. However, when use variational calculations to expand 
the exact wave function of the system 
in a basis set, only a finite number of basis functions ($N$) can be used practically. 
So, the calculated physical observable (i.e. $E_\lambda$ in this case) depends on $N$. Thus, for each $N$, the calculated energy level is denoted by $E_\lambda^{(N)}$. FSS assumes the existence of a scaling function $F_E$ such that
\begin{equation}
	\label{ScalingFun}
	E_\lambda^{(N)} = E_\lambda F_E \left( N \left| \lambda - \lambda_c \right|^\nu \right)
\end{equation}
where $\nu$ is the scaling exponent for the correlation length.
 To obtain the numerical values of the critical parameters $(\lambda_c, \alpha)$ for the  energy,
we define for any given operator ${\cal O}$ the function

\begin{equation}
\triangle_{\cal O}(\lambda;N,N')=\frac{\ln\left(\left<{\cal O}_\lambda^{N} \right>/
\left<{\cal O}\right>_\lambda^{N'}\right)} {\ln(N'/N)},
\label{fourteen}
\end{equation}

If we take the operator ${\cal O}$ to be $H-E_{th}$ and  $\partial H/\partial \lambda$, we can obtain the critical  parameters from the following function \cite{Kais-Serra}
 
\begin{equation}
\Gamma_\alpha(\lambda,N,N')=\frac{\triangle_H(\lambda;N,N')}{\triangle_H(\lambda;N,N')-\triangle_{\frac{\partial
      H}{\partial \lambda}}(\lambda;N,N')},
\label{gammafunc}
\end{equation}
which at the critical point is independent of $N$ and $N'$ and takes
the value of $\alpha$. Namely, for $\lambda=\lambda_c$ and any
values of $N$ and $N'$ we have
\begin{equation}
\Gamma_\alpha(\lambda_c,N,N')=\alpha.
\end{equation}

\noindent Because our results are asymptotic for large values of $N$, we obtain a sequence of 
pseudocritical parameters $(\lambda_N,\,\alpha_N)$ that converge to $(\lambda_c,\,\alpha)$
for $N \rightarrow \infty$. 

\subsection{Spectral methods and the exponential basis sets}

Meshfree SM is a special family of the weighted residual methods \cite{B01,C01,G01,FH03}. In these methods, the unknown functions are approximated by either an expansion of or interpolation (known as collocation method) using preselected basis sets. For homogeneous and smooth computational windows, SMs work very well. But, they suffer from the Gibbs phenomenon if any of the structural functions of the studied problem is not analytical. To avoid this problem, the computational window is divided into homogeneous domains where the discontinuities lie at the boundaries. This approach is known as multi domain spectral method (MDSM) \cite{B01,C01,G01,FH03,FH04}. In general MDSM methods allows handling very complicated and discontinuous functions. This capability is very flexible as any expansion basis set can be used. In this paper, the studied problem has a smooth structural function (i.e. Hulthen potential). So, MDSM is not used.

In many physical problems, the extensions toward infinities decay exponentially as
\begin{equation}
	\label{InfExt}
	f(x) \propto e^{\pm \beta x}
\end{equation}
where $\pm$ is used to cover both $\mp \infty$ with positive $\beta$. As aforementioned, this is one of the main challenges in SM \cite{F02,C02,G02,G03,V01,K02,S04}. A review paper by Shen and Wang discusses this problem in further details \cite{S04}. Recently, a non-orthogonal predefined exponential basis set for eignevalue problems involving half bounded domains was reintroduced \cite{FH01,FH02}. Similar sets were introduced in 1970s by Raffenetti, Bardo, and Ruedenberg \cite{B02,B03,R01} for self-consistent field wavefunctions.

The set is easy to use and it overcomes many challenges such as zero-crossing and single scaling problems by approximating the decaying domain functions by exponential basis set which spans wide range of decaying rates as follows:
\begin{equation}
	\label{ExpExpForm}
	f(x) = \sum_{n=1}^N c_n u_n (x) = \sum_{n=1}^N c_n e^{-\beta_n x}
\end{equation}
where $c_n$ are the expansion coefficients and $\beta_n$ are the pre-selected decaying rates. They are chosen intuitively based on the studied problem. But, they should allow many possible decay rates with very small number of bases. In this paper, the decaying rates are defined as
\begin{equation}
	\label{BetaForm}
	\beta_n = 10^{p_n}
\end{equation}
\begin{equation}
	p_n = d_s+\frac{n-1}{N-1} \left( d_e - d_s \right)
\end{equation}
where $d_s$ and $d_e$ are the smallest and largest used powers respectively and $N$ is the number of the used bases. 

In this paper, the set is modified slightly to have a faster convergence by enforcing the states to vanish at $x=0$. The modified set is
\begin{equation}
\label{ExpExpFormMod}
	f(x) = \sum_{n=1}^N c_n u_n (x) = \sum_{n=1}^N c_n \ x \ e^{-\beta_n x}
\end{equation}
where $\beta_n$ is as defined above in Eq. \ref{BetaForm}.

\section{Implementation}

\subsection{Formulation}

To simplify the moments calculations, the normalized Schr\"{o}dinger radial differential equation (Eq. \ref{RWE1}) is rewritten as
\begin{equation}
	\label{RWE2}
	\left[ - \frac{1}{2} \left( 1-e^{-x} \right) \frac{d^2}{dx^2} - \lambda e^{-x} \right] \psi = a^2 E_{\lambda,a} \left( 1-e^{-x} \right) \psi
\end{equation}
The expansion form (Eq. \ref{ExpExpFormMod}) is used to solve the above equation. For each used number of basis $N$, the expansion form is rewritten as follow:
\begin{equation}
	\label{ExpExpFormN}
	\psi^{(N)}(x) = \sum_{n=1}^N c_n^{(N)} \ x \ e^{-\beta_n^{(N)} x}
\end{equation}
This form is working only for bounded states and hence should work fine only for $\lambda > 0.5$. By implementing this expansion form, 
Eq. \ref{RWE2} can be written as:

\begin{equation}
	\label{SRWE01M}
	\left( \textbf{A}_N + \lambda \textbf{B}_N \right) \textbf{c}_\lambda^{(N)} = a^2 E_{\lambda,a}^{(N)} \enspace \textbf{O}_N \textbf{c}_\lambda^{(N)}
\end{equation}
where the elements of the matrices are the following scalar products:
\begin{widetext}
\begin{multline}
\label{AMElements}
\left[ \textbf{A}_N \right]_{mn} = \left< e^{-\beta_m^{(N)} x} \left| - \frac{\left( 1-e^{-x} \right)}{2x}  \left( (\beta_n^{(N)})^2 x - 2 \beta_n^{(N)} \right) - \lambda e^{-x} \right| e^{-\beta_n^{(N)} x} \right>_{1D} \\
= \frac{-(\beta_n^{(N)})^2}{2} \left( \frac{1}{\left(\beta_m^{(N)}+\beta_n^{(N)}\right)^2} - \frac{1}{\left(\beta_m^{(N)}+\beta_n^{(N)}+1\right)^2} \right) + \beta_n^{(N)} \left( \frac{1}{\beta_m^{(N)}+\beta_n^{(N)}} - \frac{1}{\beta_m^{(N)}+\beta_n^{(N)}+1} \right)
\end{multline}
\end{widetext}
\begin{multline}
\label{BMElements}
\left[ \textbf{B}_N \right]_{mn} = \left< e^{-\beta_m^{(N)} x} \left| \frac{-e^{-x}}{x} \right| e^{-\beta_n^{(N)} x} \right>_{1D} \\
= \frac{-1}{\left( \beta_m^{(N)}+\beta_n^{(N)} \right)^2}
\end{multline}
\begin{multline}
\label{OMElements}
\left[ \textbf{O}_N \right]_{mn} = \left< e^{-\beta_m^{(N)} x} \left| \frac{\left( 1-e^{-x} \right)}{x} \right| e^{-\beta_n^{(N)} x} \right>_{1D} \\
= \frac{1}{\left( \beta_m^{(N)}+\beta_n^{(N)} \right)^2} - \frac{1}{\left( \beta_m^{(N)}+\beta_n^{(N)}+1 \right)^2}
\end{multline}
In the above three equations, the integrations are taking place in one dimension and not over the physical three dimensional space. Eq. \ref{SRWE01M} is a direct eigenvalue problem and by selecting proper values for $d_s$ and $d_e$, a wide range of eigenvalues ($E_{\lambda,a}^{(N)}$) and their corresponding eigenstates ($\textbf{c}_\lambda^{(N)}$) can be calculated directly. The used values for $d_s$ and $d_e$ are -4 and 4 respectively. 

In this paper, we focus on the critical change in the lowest energy level. So in the remaining of this paper, $E_{\lambda,a}^{(N)}$ is corresponding to the calculated ground state level with $N$ basis; clearly, it is a function for $\lambda$ and the scaling factor $a$. Also, $\textbf{c}_\lambda^{(N)}$ is corresponding to the ground state and it contains the expansion coefficients. Generally, the states need normalization by dividing the coefficients by $N_{f,\lambda}^{(N)}$, where
\begin{multline}
	\label{NormalizationFactor}
\left( N_{f,\lambda}^{(N)} \right)^2 = 4 \pi \sum_{mn} c_m^{(N)*} c_n^{(N)} \int_0^\infty x^4 e^{-\left( \beta_m^{(N)} + \beta_n^{(N)} \right) x} dx \\
= 4 \pi \sum_{mn} c_m^{(N)*} c_n^{(N)} \frac{2}{ \left( \beta_m^{(N)} + \beta_n^{(N)} \right)^5}
\end{multline}
In this case and the following calculations for the potential energy, the integrations are calculated over the physical three dimensional space for the case of $l= 0$.

To apply FSS as shown later, we need to calculate the potential energy. It is simply
\begin{equation}
	\label{PotEnergy}
	V_\lambda^{(N)} = - 4 \pi \ \lambda \sum_{mn} c_m^{(N)*} c_n^{(N)} \int_0^\infty x^4 \frac{e^{-\left( \beta_m^{(N)} + \beta_n^{(N)} +1 \right) x}}{1-e^{-x}} dx
\end{equation}
The integrations are computed numerically by Gaussian quadrature. Obviously, this is the most numerically expensive part in work. However, it is clear also that the integrations are independent of the state distinctive parameters (i.e. $\lambda$ and $c_n^{(N)}$). So, for each $N$, the integrations are calculated at the beginning and the results are used to calculate $V_\lambda^{(N)}$ while varying $\lambda$.

To obtain the critical parameters, we use the  following shifted functions:
\begin{equation}
	\label{DoE}
	\Delta_E \left(\lambda; N, N', N'' \right) = \frac{ \ln \left( \dfrac{E_{\lambda,a}^{(N'')} - E_{\lambda,a}^{(N')}}{E_{\lambda,a}^{(N')} - E_{\lambda,a}^{(N)}} \right)}{ \ln \left( N'/N \right)}
\end{equation}
and
\begin{equation}
	\label{DoV}
	\Delta_{\frac{\partial
      H}{\partial \lambda}} \left(\lambda; N, N', N'' \right) = \frac{ \ln \left( \dfrac{V_{\lambda,a}^{(N'')} - V_{\lambda,a}^{(N')}}{V_{\lambda,a}^{(N')} - V_{\lambda,a}^{(N)}} \right)}{ \ln \left( N'/N \right)}
\end{equation}
 The critical parameters $\lambda_c$ and $\alpha$  can be obtained from the $\Gamma_\alpha$ as defined in Eq. \ref{gammafunc}.

\subsection{Results and discussion}

In the calculations, the scaling parameter ($a$) is set to one. Also as aforementioned, the used parameters for the exponential basis set are -4 and 4 for $d_s$ and $d_e$ respectively. These parameters are chosen after few iterations to have a reasonable accuracy for the eigenvalues. To implement FSS, $N$ is varied between 32 and 48 in a step of 2. So, $\lambda_c$ and $\alpha$ can be obtained by seeking the crossing of the FSS curves.

\begin{figure} [h]
\includegraphics[width=0.45\textwidth,height=0.35\textwidth]{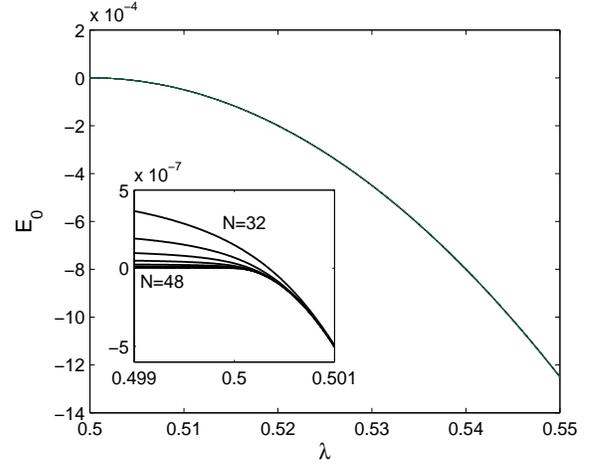}
\caption{The calculated ground state energy ($E_0$) as a function of $\lambda$ using different numbers of bases, which are varied between 32 to 48 in steps of 2. The errors are very small (as shown in Fig. \ref{ExEigenvaluesRE}) and hence the lines are overlapping and thus more resolution about $\lambda_c$ is shown in the small box.}
\label{ExEigenvalues}
\end{figure}

\begin{figure} [h]
\includegraphics[width=0.45\textwidth,height=0.35\textwidth]{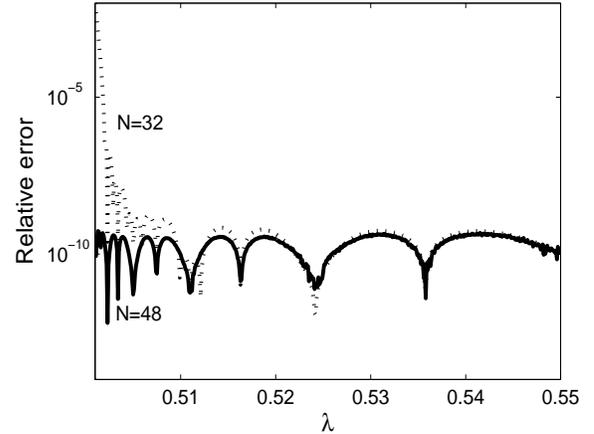}
\caption{The relative errors of the calculated ground state energy as a function of $\lambda$ for $N=32$ and $N=48$.}
\label{ExEigenvaluesRE}
\end{figure}

The calculated ground state energies ($E_0$) are shown in Fig. \ref{ExEigenvalues} as a function of $\lambda$ for all the used values of $N$. The errors are very small (as shown in Fig. \ref{ExEigenvaluesRE}) and hence the lines are overlapping. More resolution (in $\lambda$) is shown in the small box. As can be observed, the calculated values for the ground state energy start diverging slightly from the exact solutions as $\lambda$ approaches $\lambda_c$. This is expected as the used basis works for bounded states and the error shall increase as the states get extended in space. However, the calculated values of $E_0$ are still very accurate and a relative error of about $10^{-10}$ was obtained around $\lambda=0.51$ for $N=32$ and $\lambda=0.5001$ for $N=48$ as shown in Fig. \ref{ExEigenvaluesRE}.

\begin{figure} [h]
\includegraphics[width=0.45\textwidth,height=0.35\textwidth]{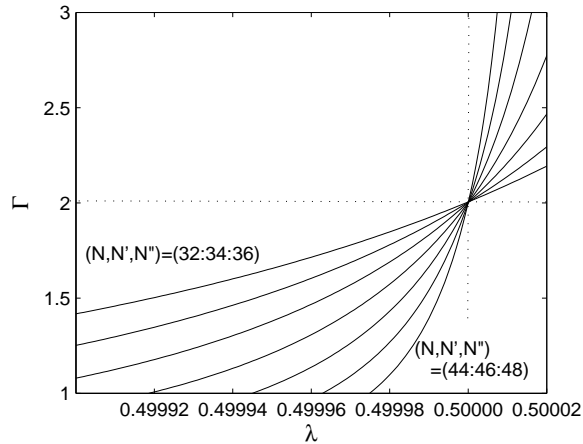}
\caption{$\Gamma_\alpha$ as a function of $\lambda$. The numbers of bases are varied between 32 to 48 in steps of 2.}
\label{GammaCrossing}
\end{figure}

In Fig. \ref{GammaCrossing}, the results of FSS calculations are shown. Plotting $\Gamma_\alpha$ as a function of $\lambda$ for different values of $N$ gives a family of curves that intersect around the analytical $\lambda_c  =0.5$ and $\alpha = 2$. The exact crossing of any adjacent curves defines the pseudo-critical parameters $\lambda_N$ and $\alpha_N$, which are used to analyse the convergence. 

\begin{figure}
\includegraphics[width=0.45\textwidth,height=0.35\textwidth]{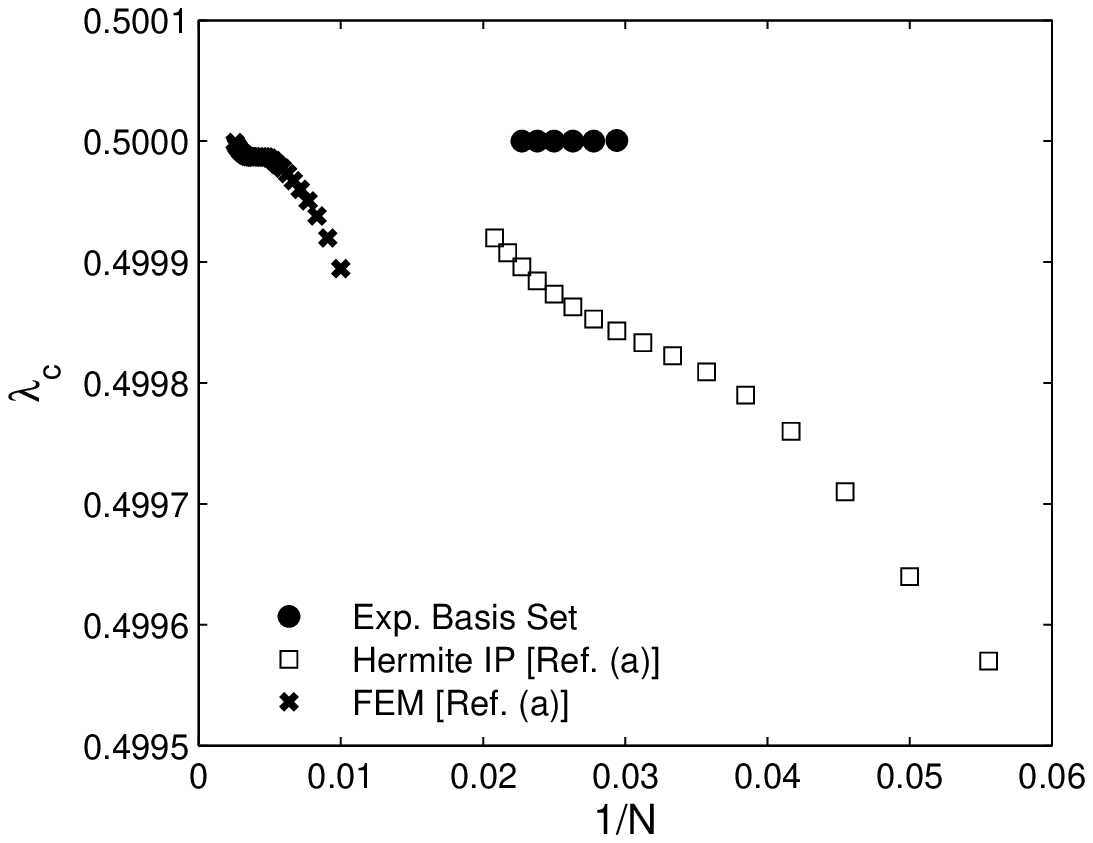}
\caption{The extrapolated values of $\lambda_c$ vs. $1/N$ as obtained by this work (solid circles), Hermite interpolation polynomials (HIP) (white squares \cite{A01}), and finite element method (FEM) (crosses \cite{A01}).}
\label{lambdaCriConv}
\end{figure}

\begin{figure} [h]
\includegraphics[width=0.45\textwidth,height=0.35\textwidth]{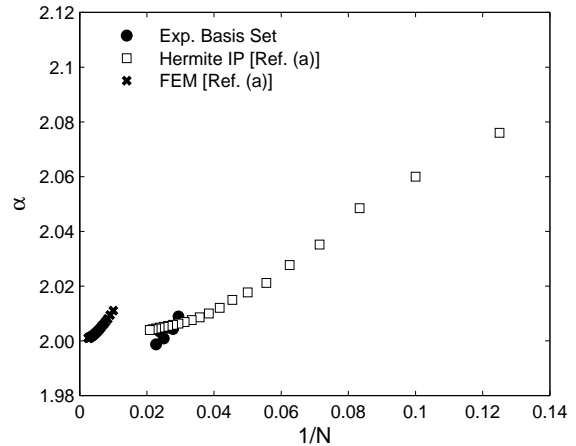}
\caption{The extrapolated values of $\alpha$ vs. $1/N$ as obtained by this work (solid circles), Hermite interpolation polynomials (HIP) (white squares \cite{A01}), and finite element method (FEM) (crosses \cite{A01}).}
\label{alphaConv}
\end{figure}

To check the convergence, the pseudo-critical parameters $\lambda_N$ (Fig. \ref{lambdaCriConv}) and $\alpha_N$ (Fig. \ref{alphaConv}) are plotted as functions of $1/N$ and compared with the results obtained using Hermite interpolation polynomials (HIP) FEM by Antillon et al. \cite{A01}. It is clear that the three methods converges to the analytical values. However, the used exponential basis set in this paper results in considerably faster convergence when compared with the other two methods. The results of the three methods are summarized in Table I.

\begin{table}
\label{TTT1}
\caption{Results for critical parameters.}
\begin{ruledtabular}
\begin{tabular}{ccccc}

	& Analytical & This work & FEM \cite{A01} & HIP \cite{A01}  \\
	
\hline

$\lambda_c$ & 0.5 & 0.500001 & 0.50184 & 0.50000 \\

$\alpha$	& 2	  & 2.00094  & 1.99993 & 2.00011 \\

$\nu$		& 1	  & 1.00000  & 1.00079 & 1.00032 
\end{tabular}
\end{ruledtabular}
\end{table}

\begin{figure}
\includegraphics[width=0.45\textwidth,height=0.35\textwidth]{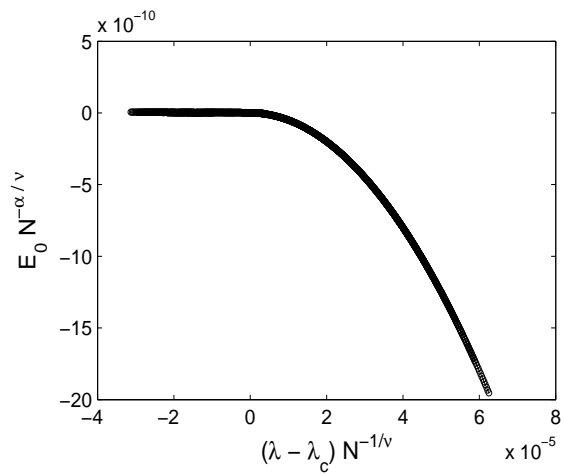}
\caption{Data collapse study of the used approach using different numbers of bases, which are varied between 32 to 48 in steps of 2.}
\label{DataCollapse}
\end{figure}

The last point to be presented is to confirm the validity of FSS assumptions using data collapse calculation. In Fig. \ref{DataCollapse}, $E_0 N^{-\alpha/\nu}$ is plotted as a function of $\left( \lambda - \lambda_c \right) N^{-1/\nu}$ for all the used $N$ values. It is clear that all the curves overlap perfectly and thus validates our FSS assumptions.

\section{Conclusion}

In atomic and molecular physics, the near threshold binding is important in the study of ionization of atoms and molecules, molecular dissociation and scattering collisions. Our benchmark calculations for the near threshold behavior of the energy levels of the Hultthen potential indicate the validity of combining FSS method with the meshfree SMs to calculate quantum critical parameters. Fortunately, the exponential basis sets  used in this study have been used previously as exponential-type even-tempered basis for atomic orbitals \cite{R01,B02,B03}. The results indicate that even-tempered bases are very accurate in Hartree-Fock  atomic calculations. Also, a systematic approach extending even-tempered atomic orbitals to optimal even-tempered Gaussian primitives have been developed and used decades ago in standard quantum chemistry calculations for atomic and molecular system \cite{Dunning,Feller,B02,B03}.  Thus, our combined FSS method and SMs based on even-tempered basis sets might be used to extract quantum critical parameters for atomic and molecular systems. In future studies, we plan to combine our FSS procedure with the Hartree-Fock and density functional theory (DFT) and other "ab initio" methods using SMs with even-tempered basis and other intuitive basis sets to analyse criticality and near threshold phenomena for molecular and extended systems. The presented approach allows scaling to analyze large systems

\section*{References}
\bibliographystyle{apsrev4-1}
\bibliography{FSSSMBibV02}

\end{document}